\documentclass{aa}  

\usepackage{graphicx}

\usepackage{url}
\usepackage[breaklinks,colorlinks,citecolor=blue]{hyperref}
\usepackage{txfonts}
\begin{document}

   \title{Chemical Inhomogeneities Amongst First Population Stars in Globular Clusters: Evidence for He variations}


   \author{C. Lardo\inst{1}
   \and
   M. Salaris\inst{2}
   \and
   N. Bastian\inst{2}
   \and
   A. Mucciarelli\inst{3}
   \and 
   E. Dalessandro\inst{4}
   \and
   I. Cabrera-Ziri\inst{5}\fnmsep\thanks{Hubble Fellow}}

   \institute{Laboratoire d' Astrophysique, Ecole Polytechnique F\'ed\'erale de Lausanne, Observatoire de Sauverny, CH-1290 Versoix, CH\\
\email{carmela.lardo@epfl.ch}
\and
Astrophysics Research Institute, Liverpool John Moores University, 146 Brownlow Hill, Liverpool, L3 		5RF, UK\\
\and
Dipartimento di Fisica e Astronomia, Universit\`a degli Studi di Bologna, via Piero Gobetti 93/2, 40129, Bologna, IT\\ 
\and
INAF-Osservatorio di Astrofisica e Scienza dello Spazio di Bologna, Via Gobetti 93/3, I-40129 Bologna, IT\\
\and
Harvard-Smithsonian Center for Astrophysics, 60 Garden Street, Cambridge, MA 02138, USA
 }

   \date{Received March XX, 2018; accepted XXX XX, 2018}

 
  \abstract{Spreads in light element abundances among stars (a.k.a. multiple populations)  
are observed in nearly all globular clusters.
One way to map such chemical variations using high-precision photometry is to employ a suitable combination
of stellar magnitudes in the F275W, F336W, F438W, and F814W filters (the so called ``chromosome map''), to maximise the separation between the different multiple populations.
For each individual cluster its chromosome map separates the so-called first population --with metal abundance patterns typical of field halo stars--  
from the second population, that displays distinctive abundance variations among a specific group of light-elements. 
Surprisingly, the distribution of first population stars in chromosome maps of several --but not all--
clusters has been found to be more extended than expected from purely observational errors, suggesting 
a chemically inhomogeneous origin. We consider here three clusters
with similar metallicity ([Fe/H]$\sim -$1.3) and different chromosome maps, namely NGC~288, M~3 and NGC~2808, and argue that the first population extended distribution (as observed in two of these clusters) is due to spreads of the initial helium abundance and possibly a small range of nitrogen abundances as well. 
The presence of a range of initial He and N abundances amongst stars traditionally
thought to have homogeneous composition, plus the fact that these spreads appear only in some clusters, 
challenge the scenarios put forward so far to explain the multiple population phenomenon.}


\titlerunning{He variations in P1 Stars}
\authorrunning{Lardo et al.}
   \maketitle
%
\section{Introduction}\label{introduzione}
Globular clusters (GCs) host multiple populations (MPs) of stars, characterised by  anti-cor\-re\-la\-tions among C, N, O, Na, and He star-to-star differences
\citep[e.g.,][]{review}.
Most scenarios for MPs invoke subsequent episodes of star formation \citep[e.g.][]{dercole08,decressin08} where stars with CNONa (and He) abundances similar 
to those observed in the 
field are the first stars to form, while stars enriched in N and Na (and He) and depleted in C and O were formed several $10^6$ up to $\sim10^8$ years 
later, from freshly synthesised
material ejected by some class of massive stars 
from the first epoch of star formation.
In the following, we denote as first population (P1), stars with field-like light element pattern. We refer to stars with enhanced
N and Na as second population (P2).
However, all the existing models have difficulties in quantitatively reproducing
observations, and no consensus has been reached on the mechanism responsible for MP origin \citep[e.g.][]{larsen14,kruijssen15,nateHE,bastian15,renzini15,review}.

Key constraints on MPs can be derived from photometry of individual GC stars when 
suitable filters are used. Recently, the Hubble Space Telescope (HST) UV legacy survey of Galactic GCs \citep[e.g.,][]{piotto15,soto17} has
provided accurate photometry for 57 Galactic GCs 
in the Wide Field Camera 3 (WFC3) filters F275W, F336W, and F438W on board HST, perfectly suited to undertake photometric studies of MPs.
Indeed, filters covering wavelengths shorter than $\lesssim$4500~\AA~are particularly sensitive to star-to-star differences in C, N, and O content. 
When complemented with the existing 
optical photometry from the Wide Field Channel of the HST Advanced Camera for Survey \citep[WFC/ACS;][]{sarajedini07} in the F606W and F814W filters, these UV observations 
have been  widely used to identify MPs in individual GCs 
and characterise their properties \citep[number ratio between P1 and P2 stars, radial distributions, degree of N enrichment, and trends of 
such properties with cluster parameters; e.g.][]{piotto15,milone17}. 

More specifically, \citet{milone15, milone17} introduced the pseudo two-colour diagram $\Delta$(F275W$-$F814W) vs. $\Delta$C(F275W, 
F336W, F438W) -- also called chromosome map -- where different GC sub-populations can be easily identified, especially when
considering red giant branch (RGB) stars. 
RGB P1 stars which do not show Na-O abundance variations are expected to be generally distributed around the origin of the chromosome map and cover a 
narrow range of $\Delta$C(F275W, F336W, F438W) values, whilst P2 stars (e.g., Na-rich and O-poor stars) span a wide range of both $\Delta$C(F275W, F336W, F438W) 
and $\Delta$(F275W$-$F814W) values \citep{milone15,milone17,carretta18}. 
 
However, as reported by \citet{milone17}, in the majority of the clusters the $\Delta$(F275W$-$F814W) values (and $\Delta$C(F275W, F336W, F438W) to a lesser extent) for P1 stars display a range much larger than
expected from the photometric errors, implying a chemical inhomogeneity. This  
is a puzzling result, given that according to self-enrichment models of MP formation, these stars should have all essentially the same chemical composition within 
individual clusters \citep[e.g.][]{grattonREV,renzini15}, hence also the $\Delta$(F275W$-$F814W) distribution of P1 stars should be narrow.
The analysis by \citet{milone17} points out this puzzling result, but does not investigate the origin of the extended P1 distributions.

In this paper, we use synthetic spectra to investigate the origin of the extended distribution of P1 stars in the chromosome maps, by studying 
three clusters with similar iron content ([Fe/H]$\sim -$1.3; to minimise metallicity effects on the colours of stars) and different chromosome maps, namely NGC~288, M~3 (NGC~5272) and NGC~2808. M~3 and NGC~2808 were selected because they display a well populated P1 in their chromosome maps, that spans large ranges in $\Delta$(F275W$-$F814W), as large as those covered by P2 stars. NGC~288 is included as a counter-example, given that P1 stars in this cluster have nearly homogeneous $\Delta$C(F275W, F336W, F438W) and $\Delta$(F275W$-$F814W) values.

The paper is structured as follows:
We describe the data in \S~\ref{OSSERVAZIONI}, and discuss the observed
P1 chromosome maps in terms of abundance variations in \S~\ref{CROMO}. Conclusions follow in \S~\ref{CONCLUSIONI}.


\section{The observational data-set and cluster sample}\label{OSSERVAZIONI}

Catalogs for NGC~288, M~3, and NGC~2808  
 are from the first public data release of the HST UV legacy survey of Galactic Globular Clusters and they are described in 
 \citet{soto17}.
The data used in this work is different from that in \citet{milone17} although this does not affect the main result of our investigation. The available catalogs are based on a one-pass reduction with static
library point-spread functions and contain only those 
stars that were already found in \citet{sarajedini07}. The colour-magnitude diagrams (CMDs) shown in \citet{piotto15} and \citet{milone17} are based on a different treatment of each cluster 
(including custom point spread function modelling for each exposure, UV-based star lists, etc.; see \citealp{soto17}).
A collection of chromosome maps for the entire 
sample of the UV survey from \citet{piotto15} data can be found in \citet{milone17}.

\begin{figure}
 \centering
\includegraphics[width=0.79\columnwidth]{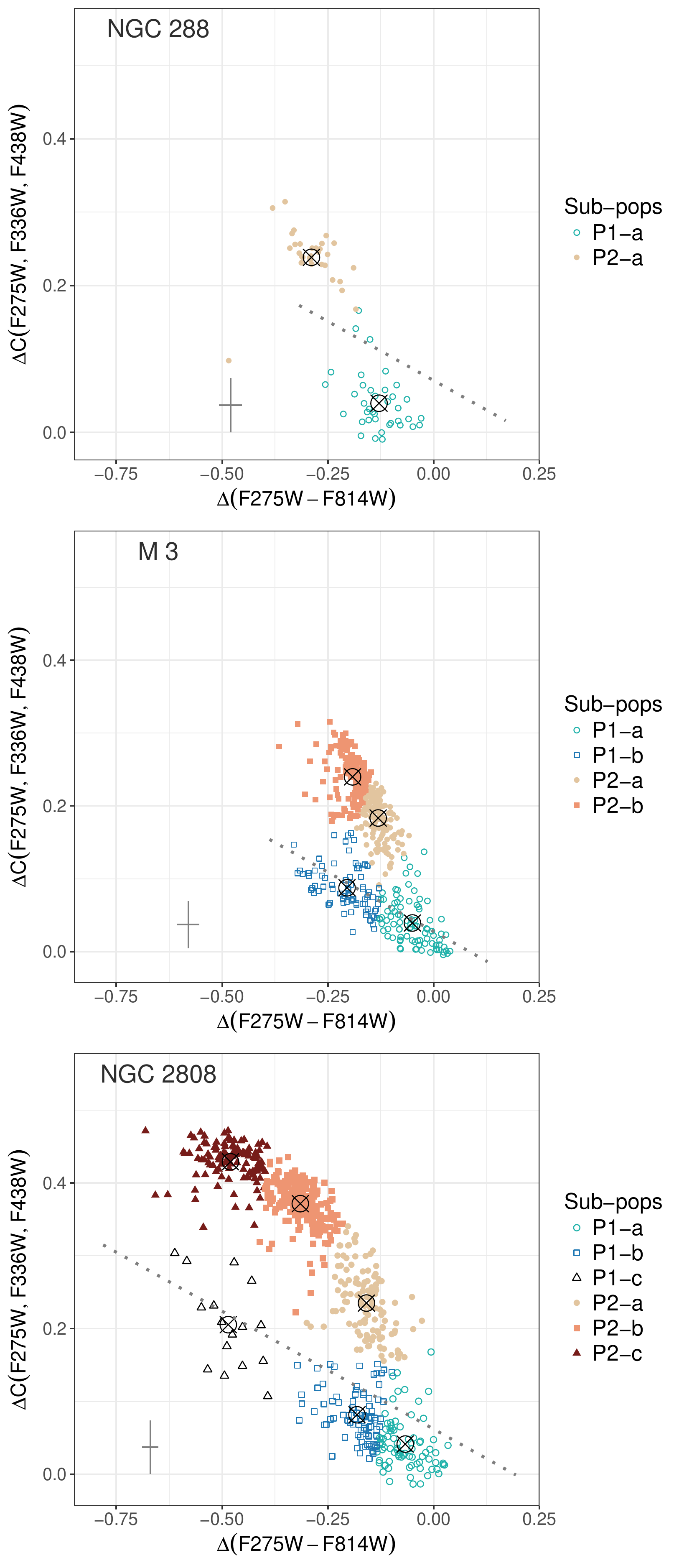}
\caption{
Chromosome maps for RGB stars in the studied clusters. 
Dotted lines separate between P1 and P2 according to the definition given in \citet{milone17}. 
The mean photometric errors are shown in the bottom left corner of each panel. 
Different sub-populations identified employing a 
$k$-means algorithm are shown with different colours and symbols, as labelled. Large empty black symbols denote the mean $\Delta$ (F275W-F814W) and $\Delta$C(F275W, F336W, F438W)  
values of each sub-population. 3D chromosome maps can be visualised at these links \href{https://carlardo.github.io/personal_site/animations/n288.html}{NGC~288}, 
\href{https://carlardo.github.io/personal_site/animations/m3.html}{M 3},  \href{https://carlardo.github.io/personal_site/animations/n2808.html}{NGC~2808}.}
\label{fig:CLUSTERING}
\end{figure}

Given that we are interested in high-precision photometry of cluster members, we retain for our analysis only stars with a displacement in pixels less 
than $\leq$ 0.1 in both the $x$- and $y$-axis.  We also limited  our study to stars with small photometric uncertainties
($<$0.03, 0.03, 0.02 in F275W, F336W, and F438W, respectively). 
RGB stars are selected as those stars lying onto the narrow RGB main locus in  the F814W-(F606W$-$F814W) CMD, where the RGB sequence broadening due to light-element 
inhomogeneities  
is minimised \citep[e.g., ][]{sbordone11}. Selected RGB stars are then plotted in the F814W-C(F275W, F336W, F438W)$\equiv$(F275W$-$F336W)$-$(F336W$-$F438) and F814W-(F275W$-$F814W) 
diagrams, to measure for each star its distance in C(F275W, F336W, F438W) and (F275W$-$F814W) with respect to reference ridge lines located at the blue and
the red side of the RGB \citep{milone17}. 
Values for each star are then normalised, respectively, to the C(F275W, F336W, F438W) and (F275W-F814W) differences between the red and blue fiducials,
taken 2.0 magnitudes above
the main sequence turnoff in the F814W filter \citep[these differences are denoted as ${\rm W_{C (F275W,F336W,F438W)}}$ and ${\rm W_{m_{F275W}-m_{F814W}}}$ in][]{milone17}.

We denote these indices for each RGB star as $\Delta$C(F275W, F336W, F438W) and $\Delta$(F275W$-$F814W), respectively, following \citet{milone17}. 
The normalisation described above implies that the range of $\Delta$(F275W$-$F814W) and $\Delta$C(F275W, F336W, F438W) covered by stars in each cluster 
will be equal to the values of ${\rm W_{C (F275W,F336W,F438W)}}$ and ${\rm W_{m_{F275W}-m_{F814W}}}$ specific to each individual cluster. 
These two quantities vary from cluster-to-cluster \citep{milone17} and depend 
on the specific chemical make-up of the underlying stellar populations.

Finally, $\Delta$C(F275W, F336W, F438W) and $\Delta$(F275W--F814W) are plotted against each other 
to derive the chromosome maps in Fig.~\ref{fig:CLUSTERING}. The same figure displays also the line used to separate between P1 and P2,
as defined by  
\citet{milone17}. P2 stars are those located above the dotted line of Fig.~\ref{fig:CLUSTERING}, while P1 stars are located below.
This separation is based on the fact that spectroscopic analyses of a subsample of stars in NGC~6121 (M~4) included in the chromosome maps 
of \citet{milone17} show Na-O ratios typical of field halo stars, for {\bf objects lying in the P1 region of the maps}.  
These photometric maps are very similar to the corresponding ones published by \citet{milone17} for the same clusters, and the values of  ${\rm W_{C (F275W,F336W,F438W)}}$ 
and ${\rm W_{m_{F275W}-m_{F814W}}}$ we obtain are consistent with the corresponding values in Table~2 of \citet{milone17}.

\begin{table*}

\setlength{\tabcolsep}{5pt}
\caption{Mean values of $\Delta$C(F275W, F336W, F438W) and $\Delta$(F275W$-$F814W) for the sub-populations of Fig.~\ref{fig:CLUSTERING}. 
Values for [Fe/H] and mass of each cluster, and the number of stars in each chromosome map sub-population are also reported. Metallicities for NGC~288 and NGC~2808 are from \citet{carretta2009}. The tabulated [Fe/H] for M~3 is from \citet{cohenM3}. The total cluster mass log M$_{\rm tot}$ is from \citet{mc05}.}
\label{TAB}

\centering
\begin{tabular} {lccccccc}
\hline\hline

Cluster & log M$_{\rm tot}$ & [Fe/H] & Sub-pop  & $\langle \Delta_{\rm (F275W-F814W)} \rangle$  & $\langle \Delta {\rm _{C(F275W, F336W, F438W)}} \rangle$ &
N$_{\rm stars}$  \\
&(M$\sun$)& (dex) &  &  &  & \\

\hline

  NGC~288 & 4.85& -1.30 & P1-a & -0.129& 0.040 &   39 \\
                   &        &          & P2-a & -0.289& 0.238 &   30  \\
\hline
        M~3   & 5.58 & -1.39 &  P1-a & -0.050 & 0.039 &  89  \\
                   &        &          & P1-b & -0.204 & 0.087 &  77  \\
                   &       &           & P2-a & -0.131 & 0.183 & 148  \\
                  &        &           & P2-b & -0.192 & 0.240 & 190  \\      
\hline
NGC~2808 & 5.93 & -1.15  & P1-a &  -0.067  &  0.041  &	83   \\
        & & & P1-b &  -0.181  &  0.082  &	88  \\ 
	& & & P1-c &  -0.485  &  0.205  &	16   \\ 
	& & & P2-a &  -0.159  &  0.235  &	128  \\ 
	& & & P2-b &  -0.315  &  0.371  &	217  \\ 
	& & & P2-c &  -0.471  &  0.420  &	116  \\      
\hline

\end{tabular}
\end{table*}

\section{The chromosome maps}\label{CROMO}

Figure~\ref{fig:CLUSTERING} shows chromosome maps for NGC~288, M~3, and NGC~2808.
Stars in each cluster display their own distinctive pattern. We resorted to a clustering analysis to identify the sub-populations shown in the figure,  
employing a $k$-means algorithm \citep[see, e.g.,][]{hartigan}.
Most notably, the $\Delta$(F275W$-$F814W) and $\Delta$C(F275W, F336W, F438W) ranges spanned by both P1 and P2 stars in M~3 and NGC~2808 are significantly 
larger than expected from photometric errors alone (see Fig.~\ref{fig:CLUSTERING}). As noted by \citet{milone15, milone17}, this means that 
both P1 and P2 groups host stars with inhomogeneous chemical compositions.
While it is well established that P2 stars include different sub-population with varying degrees of chemical enrichment \citep[e.g., ][and references therein]{carretta2009, car09b,milone15}, 
P1 stars are not expect to display any intrinsic chemical variations within the standard scenarios proposed to explain the multiple population phenomenon.
Still, the distribution of P1 stars along the $\Delta$(F275W-F814W) and $\Delta$C(F275W, F336W, F438W) axes is broader than what expected   
from observational errors, for both M~3 and NGC~2808 (and for a large fraction of the 57 clusters presented in \citealp{milone17}, as also noted by these authors).
Conversely, in case of NGC~288, only two main populations with homogeneous internal composition can be identified. In particular, P1 stars appear to be strikingly homogeneous compared to the other 
two clusters.

\begin{figure}
\centering
\includegraphics[width=0.65\columnwidth]{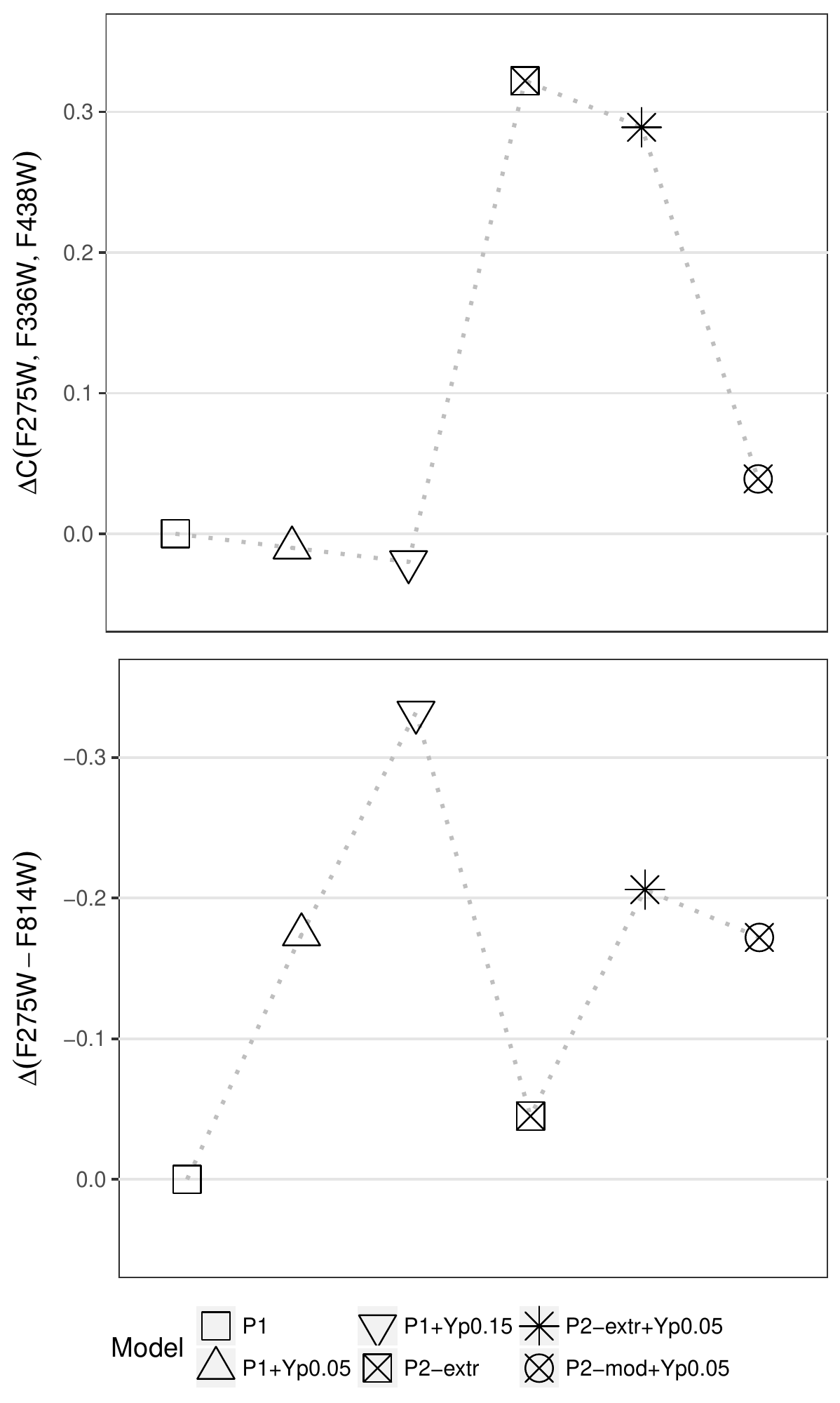}
\caption{Expected chromosome map indices of RGB stars with P1 chemical composition (squares), P1 with He enhanced by $\Delta$Y=0.05 (triangles up), P1 with He enhanced by $\Delta$Y=0.15 (triangles down), P2 extreme (crossed squares), P2 extreme with $\Delta$Y=0.05 (stars), 
and P2 moderate with $\Delta$Y=0.05  (crossed circle), --see text for details.}
\label{fig:ATLAS}
\end{figure}

\subsection{Reading chromosome maps in terms of abundance variations}\label{teoria}

As shown by, e.g. \citet{piotto15} and \citet{lagioia}, the (F275W$-$F814W) colour of RGB stars is affected by variations of the oxygen abundance due to the presence in the 
spectrum of an OH molecular band within the 
F275W filter bandpass. At low [Fe/H] the effect of a substantial decrease of the oxygen abundance by 0.3~dex on the (F275W-F814W) colour is very small if not negligible, increasing 
only for [Fe/H] around and above $\sim -$1.0~dex. 
This means that a variation of the initial oxygen cannot explain the extended P1 in the chromosome maps of clusters like M3 discussed in this paper. 
In addition, the location of P1 stars in the chromosome maps is defined --as already mentioned-- in terms of their stellar Na and O abundances. 
\citet{milone17} shows that stars with spectroscopically measured Na-O ratios typical of field halo stars --hence homogeneous in Na and O-- 
can be seen to occupy the full $\Delta$(F275W$-$F814W) range of the extended P1 in the cluster M~4 ([Fe/H]$\sim -$1.0; see also \citealp{milone15,carretta18}). 
The availability of recent Na and O abundance determinations for RGB stars in NGC~2808 by \cite{carretta18}, demonstrates  
the P1 stars as defined from the chromosome maps following \citet{milone17}, 
all have homogeneous low-Na and high-O, typical of the field halo population\footnote{\cite{carretta18} use a different pseudo-colour diagram in their analysis, but the 
correlation between the position of stars in their diagram and the chromosome maps can be easily established from the work by \cite{milone15}. These latter authors 
cross-correlated stars in both the diagram employed by \cite{carretta18} and a chromosome map for NGC~2808, even if at that time they had not employed yet the 
term {\it chromosome map}. This also means that the identification of P1 stars as stars with homogeneous field-like [Na/Fe] and [O/Fe] abundance ratios  
in terms of their location in the chromosome map of a generic cluster seems to be well established.}.

To understand the origin of the extended P1 we have proceeded as follows. It is well known since \citet{salaris06}, \citet{sbordone11} and \citet{cassisi13} that 
theoretical RGBs of P2 stars overlap with P1 RGBs in the ${\rm log(L/L_{\odot})-log(T_{\rm eff})}$ diagram, unless the initial He abundance is different.
The analysis by \citet{vdb} shows however that variations of Mg and Si at constant Fe (the clusters analysed here as well the large majority of
  Galactic GCs --bar a few exceptions-- with or without an extended P1 do not show internal inhomogeneities in Fe abundances) are able to change the $T_{\rm eff}$ of metal poor RGB models.
Potentially, a range of Si and/or Mg amongst P1 stars may produce a spread in the chromosome maps of a given cluster. This however cannot be the case for the clusters with an extended P1 studied here. Let's consider 
M~4 \citep[chromosome map shown by][]{milone17} and NGC~2808 that we analyse here. The spectroscopic analysis by \cite{Carretta:09UVES} shows that P1 stars (defined as stars with low Na and high O, 
consistent with their identification in the chromosome maps) in NGC~2808 have homogenous Mg (with typical field-like abundances) and also homogeneous Si.
Also in M~4 there is no spread of Mg and Si in P1 stars.  

This means that, at fixed Y, one can model the effects of C, N, O, Na variations by just calculating synthetic spectra with the appropriate composition, keeping 
the stellar parameters unchanged. We know also that Na variations do not 
affect the bolometric magnitudes in the filters used for the chromosome maps \citep[see, e.g.][]{sbordone11, piotto15}.

We have taken as reference parameters for our synthetic spectra calculations T$_{\rm{eff}}$ = 5260~K and surface gravity log(g)=3.3 (cgs units) corresponding to 
a RGB star along a representative [Fe/H]=$-$1.3, 12~Gyr $\alpha$-enhanced, initial Y=0.246 BaSTI isochrone \citep[][]{Pietrinferni:06,Pietrinferni:09}, taken about 2.0 magnitudes above
the main sequence turnoff in the F814W filter. This is the RGB level to which the chromosome maps' indices are normalised. 

In the spectrum calculations we have considered first a representative P1 composition with [C/Fe]=--0.3, [N/Fe]=+0.2, [O/Fe]=+0.3  \citep{cohen02,carretta2009}, all other heavy elements scaled as in the $\alpha$-enhanced metal mixture of the BaSTI models \citep[see, e.g.,][]{sbordone11}, and Y=0.246.

We have then considered a P2 composition with [C/Fe]= --0.4, [N/Fe]=+0.5, [O/Fe]=+0.2 (denoted as P2-moderate) keeping all other abundances unchanged, and a more extreme P2 composition with [C/Fe]= --1.0, [N/Fe]=+1.3, [O/Fe]=--0.3 (denoted as P2-extreme). These patterns are not necessarily characteristic of the three clusters studied here, since no C, N, and O measurements ({\em in the same star}) are available in literature, but they rather reflect the
standard trends observed in Galactic GCs at this intermediate metallicity \citep{cohen02,carretta2009}.

In addition, we calculated spectra of stars with P1, P2-moderate and P2-extreme compositions including also an enhancement of the He mass fraction, by $\Delta$ Y=0.05. A variation of He affects the T$_{\rm{eff}}$ of the RGB at the magnitude level assumed for the normalisation of the chromosome maps' pseudo-colours.
To this purpose, we have first estimated from BaSTI isochrones that this Y enhancement causes an increase of the reference RGB T$_{\rm{eff}}$ by $\sim$70~K\footnote{The corresponding change of surface gravity is small and has a negligible effect on the spectra.}.

We have then calculated spectra with the mentioned P1 and P2 metal distributions, considering a 70~K T$_{\rm{eff}}$ increase of the corresponding model atmospheres with respect to the reference value adopted for the P1 and P2 compositions with {\sl normal} Y.

Finally, we computed a spectrum for a star with P1 composition and He enhancement $\Delta$ Y=0.15 (in this case the effect of the He increase on the reference RGB T$_{\rm{eff}}$ is an increase of 170~K.

Calculations have been performed with the {\tt ATLAS12} \citep[][]{kurucz05,castelli05,sbordone07} and {\tt SYNTHE} \citep{kurucz05} codes, in the wavelength range between 2000--10600~\AA. 
Bolometric corrections in the VEGAMAG system have been calculated for the F275W, F336W, F438W filters of 
UVIS/WFC3 and the F606W, F814W filters of ACS/WFC following \citet{girardi}, to determine C(F275W, F336W, F438W) and  (F275W$-$F814W) 
differences with respect to a reference RGB star with P1 composition and standard helium. These differences correspond to values of the  
$\Delta$C(F275W, F336W, F438W) and $\Delta$ (F275W-F814W) indices for these synthetic P2 and P1 enhanced helium populations, given that the canonical P1 with standard cosmological He 
is expected located at coordinates $\Delta$C(F275W, F336W, F438W)=0.0 and  $\Delta$ (F275W-F814W)=0.0 in the chromosome maps.  

The results are plotted in the two panels of Fig.~\ref{fig:ATLAS}. We display on the horizontal axis the different chemical compositions investigated. The vertical axis 
displays the indices $\Delta$C(F275W, F336W, F438W) and $\Delta$(F275W$-$F814W) for all our synthetic populations. 
The results are quite clear. The index $\Delta$(F275W$-$F814W) is affected essentially only by variations of Y, whilst the $\Delta$C(F275W, F336W, F438W) index 
is in comparison weakly sensitive to Y, but it is mainly affected by variations in N  \citep[e.g.][]{piotto15}.
The range of $\Delta$(F275W$-$F814W) values for P1 stars as observed in several clusters like M~3 and NGC~2808 discussed here, is therefore due to a range of Y abundances. 
This is however not found in other clusters, like NGC~288.
The extended P1 sequences in the chromosome maps display also a tilt towards increasing $\Delta$C(F275W, F336W, F438W) for decreasing $\Delta$(F275W$-$F814W) \citep[e.g.][]{milone17}.
Based on Fig.~\ref{fig:ATLAS}, this can be possibly explained by a small range of N amongst P1 stars. 

We conclude that the spread in P1 stars indicates that a relatively large spread of Y is present within this population, possibly accompanied by a small range in N.  

In summary, P1 stars are defined by their location in the chromosome maps, that is consistent with the 
position of stars with homogeneous and field-like [O/Fe] and [Na/Fe] measurements in clusters like M~4, and NGC~2808 that we investigated here.
These P1 stars are often 
called first generation stars in the literature, according to the proposed cluster self-enrichment scenarios. 
We have shown here that P1 stars in clusters with extended P1 in their chromosome maps like NGC~2808 and M~3, 
show a spread of initial Y and a small range of N. In other cases, like NGC~288 with a non-extended P1 in the chromosome map,
also the Y and N abundances are homogeneous, like  O and Na.

The former behaviour is entirely unexpected in the context of all scenarios for the origin of MPs.

\begin{figure*}
  \begin{minipage}[b]{0.48\linewidth}
    \centering
    \includegraphics[width=1.0\linewidth]{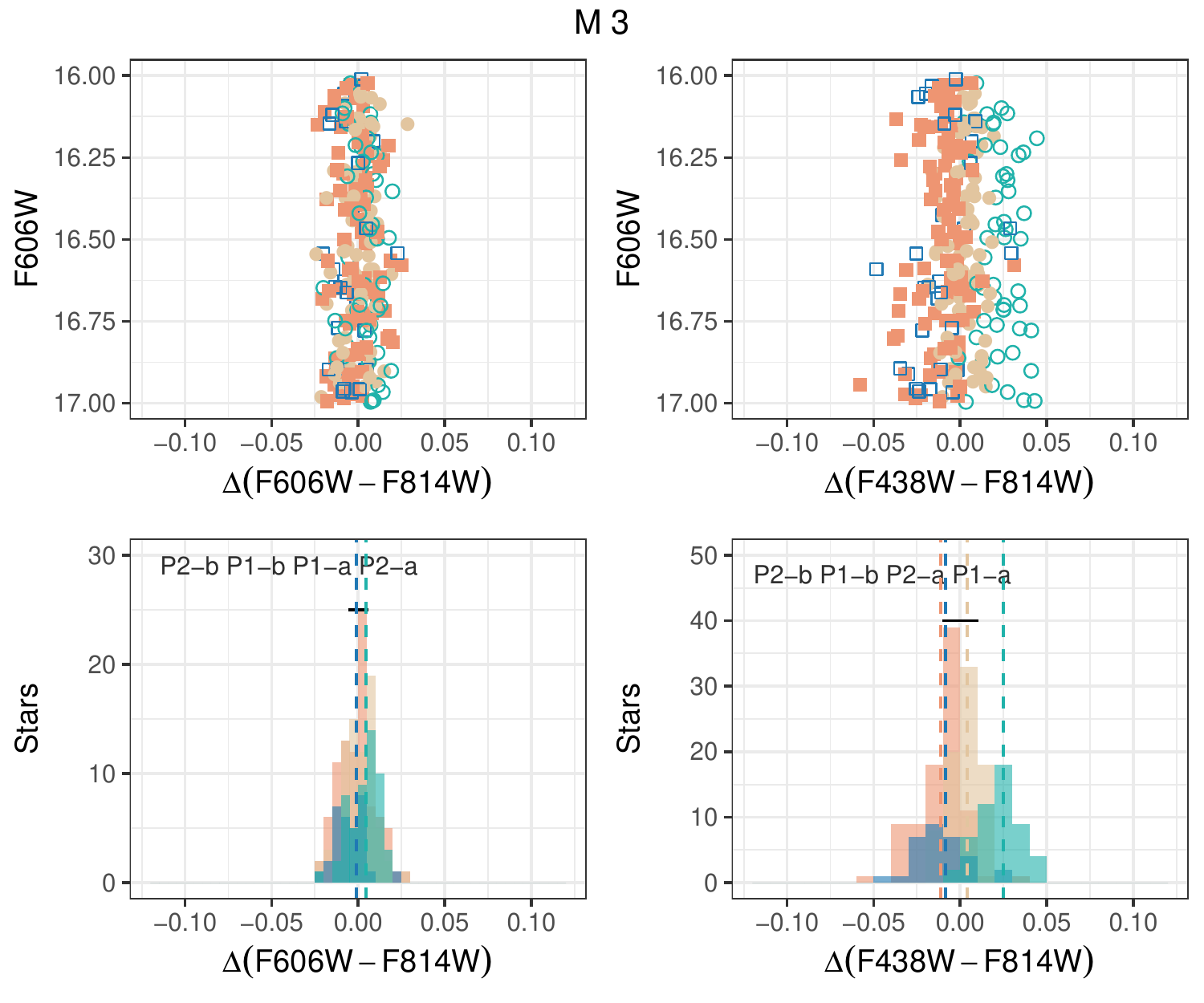}
  \end{minipage}
  \hspace{0.05cm}
  \begin{minipage}[b]{0.48\linewidth}
    \centering
    \includegraphics[width=1.0\linewidth]{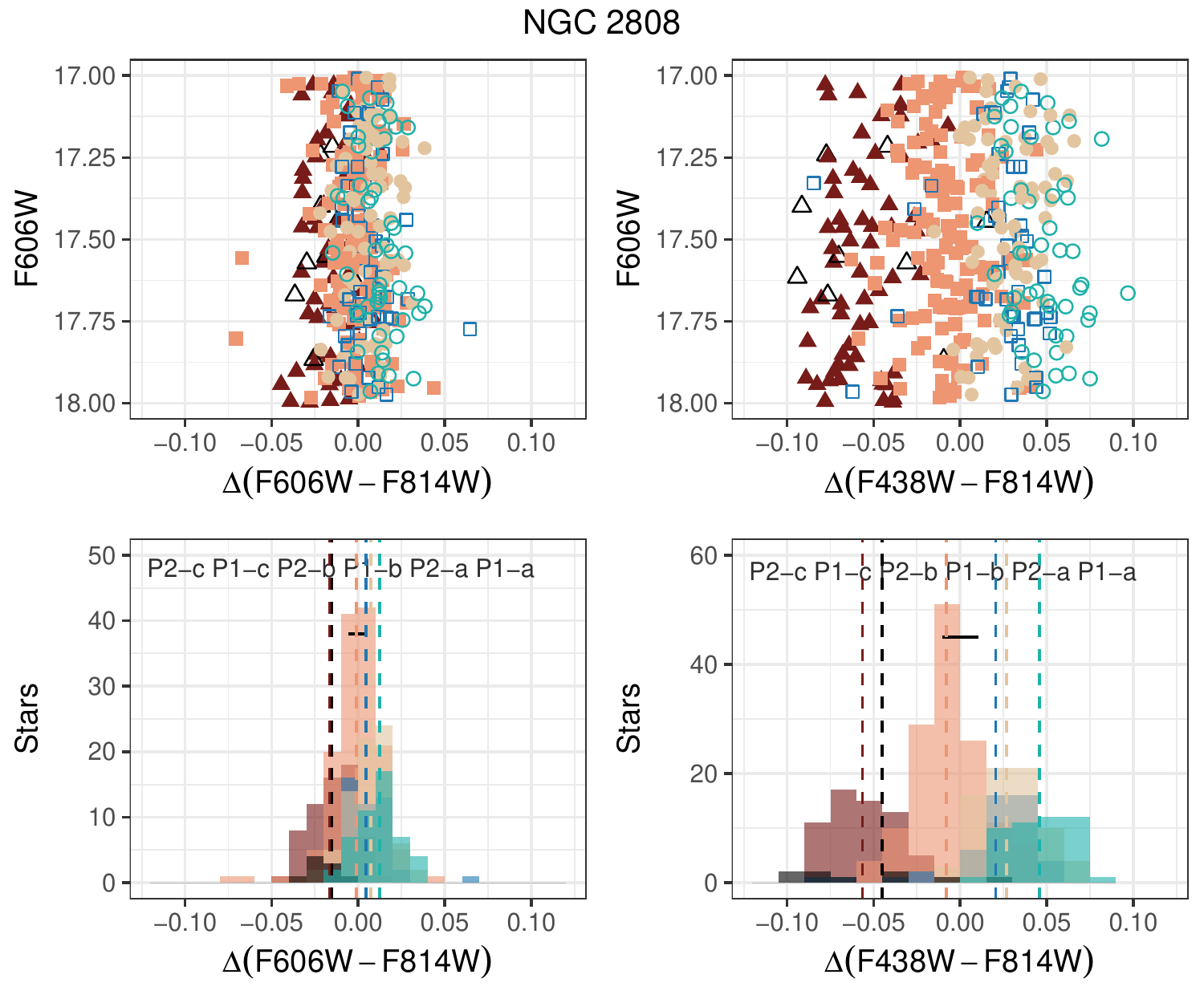}
\end{minipage}
\caption{The top panels display F606W-$\Delta$(F606W$-$F814W) and F606W-$\Delta$(F438W$-$F814W) diagrams for the various M~3 and NGC~2808 RGB subpopulations. 
Symbols are the same as in Fig.~\ref{fig:CLUSTERING}. The lower panels display histograms  
of $\Delta$(F606W$-$F814W) and $\Delta$(F438W$-$F814W) for both clusters. Dotted lines denote the mean values of these colour differences 
for each subpopulation (see text for details).}
\label{fig:BVI}
\end{figure*}

\subsection{The distribution of the sub-populations along the RGB }

An independent test for He-abundance variations among P1 stars is to look 
at the optical colour distributions of each of the identified sub-populations.  
The reason is that bolometric corrections in optical filters  
are negligibly affected when moving from typical P1 to P2 metal distributions \citep[e.g.][]{sbordone11}, but 
an increase of Y shifts the RGB towards bluer colours because isochrones with enhanced He have hotter T$_{\rm{eff}}$ values. 

\begin{figure*}
 \begin{minipage}[b]{0.48\linewidth}
    \centering
    \includegraphics[width=\linewidth]{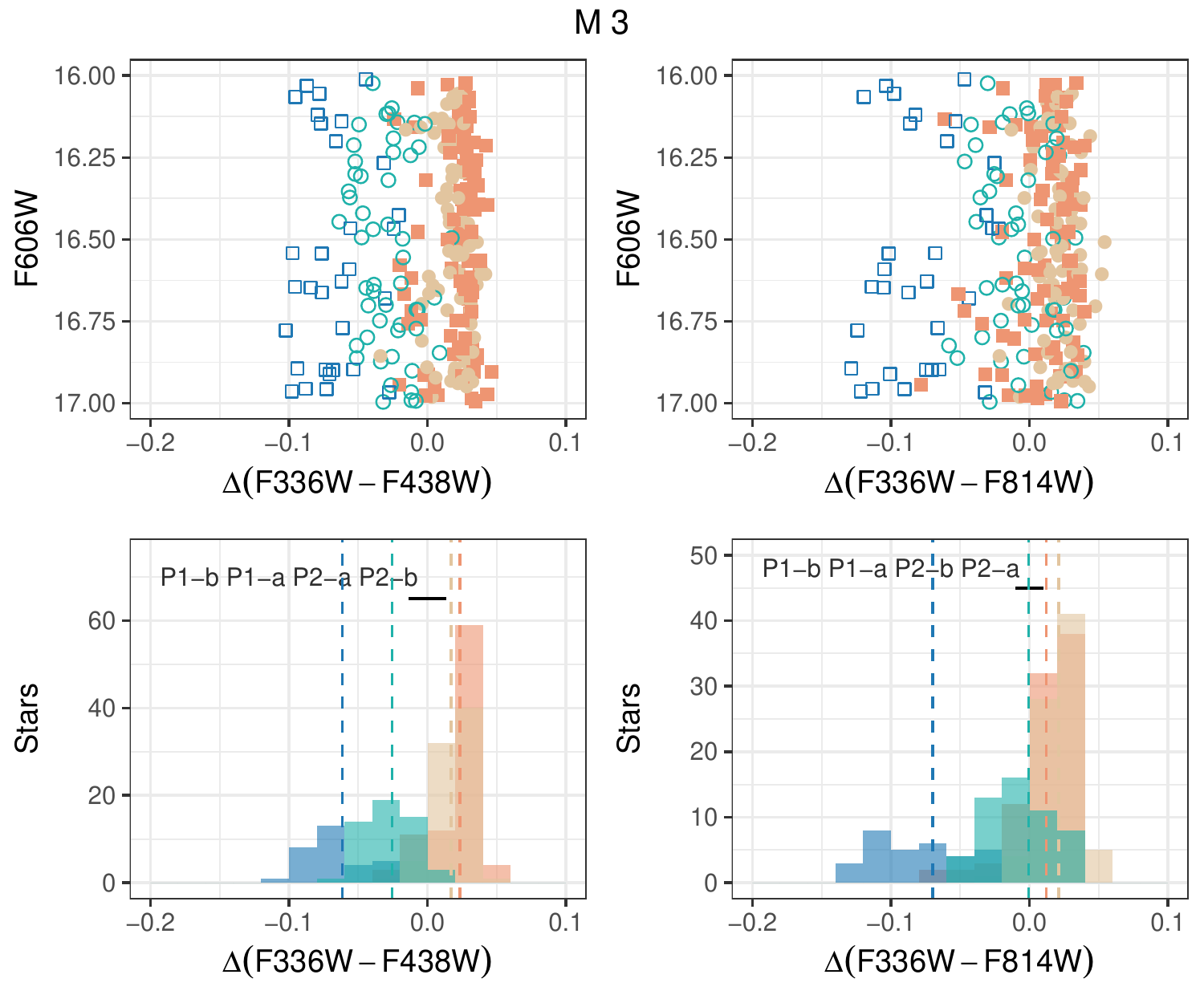}
  \end{minipage}
  \hspace{0.05cm}
  \begin{minipage}[b]{0.48\linewidth}
    \centering
    \includegraphics[width=\linewidth]{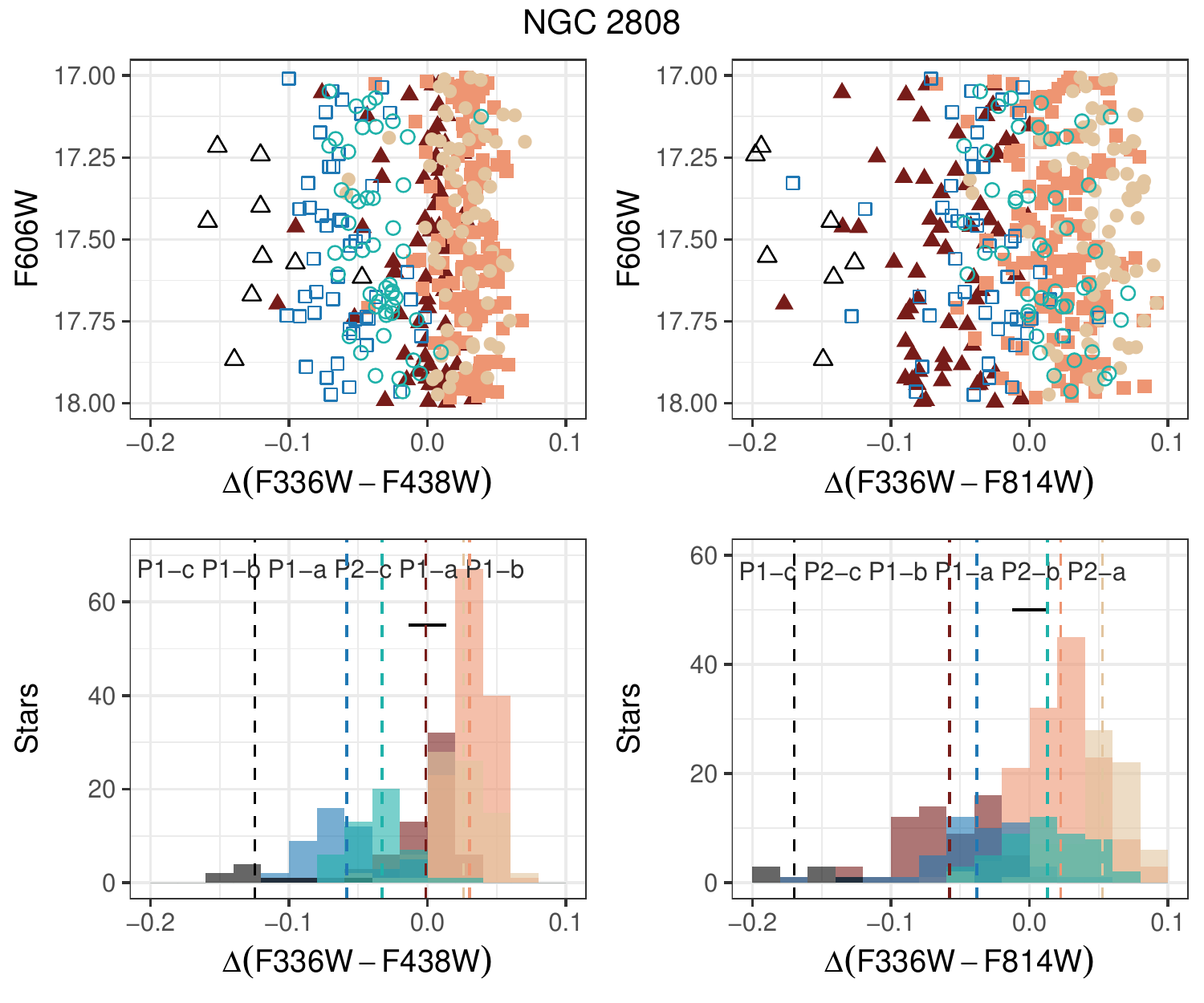}
\end{minipage}

\caption{The same as Fig.~\ref{fig:BVI} but for $\Delta$(F336W-F438W) and $\Delta$(F336W-F814W) colour differences (see text for details).}
\label{fig:UBV}
\end{figure*}

For each cluster, we have first identified different RGB sub-populations from the chromosome maps of Fig.~\ref{fig:CLUSTERING}. 
The mean values of $\Delta$C(F275W, F336W, F438W) and $\Delta$ (F275W-F814W) indices, plus the number of stars belonging to each sub-population are in Table~\ref{TAB}. 
The P1 population in M3 appears to be made of two subpopulations denoted as P1-a and P1-b, when moving from $\Delta$ (F275W-F814W)=0.0 towards negative values.
NGC~2808 displays three P1 subpopulations, denotes as P1-a, P1-b, P1-c. Both clusters display a number of P2 subpopulations, two in case of M~3 and three in case 
of NGC~2808. 
We have then considered the F606W-(F606W$-$F814W) and F606W-(F438W$-$F814W) CMDs, including RGB stars from one magnitude below the RGB bump 
up to the bump location, a well defined and populated  
part of the RGB. A RGB fiducial line using a polynomial regression was determined in each CMD, and for each star the colour differences $\Delta$(F606W-F814W) and $\Delta$(F438W-F814W) 
with respect to the colour of the fiducial line at the same F606W magnitude have been determined. 
Figure~\ref{fig:BVI} displays the resulting F606W-$\Delta$(F606W$-$F814W) and F606W-$\Delta$(F438W$-$F814W) diagrams, plus the histograms of these colour differences 
for both clusters.

The larger baseline (F438W$-$F814W) colour is more sensitive to the stellar ${\rm T_{eff}}$ (e.g., 
${\rm T_{eff}}$ changes are proxy for He variations), and in the F606W$-$(F438W$-$F814W) CMDs of both clusters P1 stars 
become on average increasingly bluer when moving from subpopulations P1-a to P1-b in case of M~3, and P1-a to P1-c in case of NGC~2808. This is consistent with the interpretation that
P1 stars encompass a range of initial He abundances. The separation is larger for NGC2808, consistent with a larger Y difference amongst the three P1 subpopulations relative to M~3.
A separation can be seen also in the (F606W$-$F814W) colour --less sensitive to ${\rm T_{eff}}$-- for NGC2808, but not for M~3, consistent again with a larger He range for NGC2808 P1 stars.
Making use of BaSTI isochrones with varying initial Y, 
the mean $\Delta$(F606W$-$F814W) colour differences between P1-a and P1-b in M~3, and  P1-a  and P1-c in NGC~2808 translate into $\Delta$Y$\sim$ 0.024 and $\sim$0.125, respectively.

As a consistency check we analysed in the same way the F606W - (F336W--F438W) and 
F606W - (F336W--F814W) CMDs, that are sensitive also to the different P1 and P2 metal mixtures 
(see Fig.~\ref{fig:UBV}).
In the previous optical CMDs, P2 stars in both clusters were shifted to the blue compared to the P1-a --with lower He-- subpopulation, whereas at these shorter wavelengths the effect of the 
metal mixture is expected to shift P2 stars towards the red compared to P1-a stars, counterbalancing or even reversing the effect of an increased He. 
On the other hand, P1 stars with increased He should still be shifted to the blue compared to P1-a objects, despite the small enhancement of N necessary to explain their chromosome maps.
This is exactly what we find, when comparing the mean colour differences of the various subpopulations with respect to the fiducial lines, as shown 
in the lower panels of Figs.~\ref{fig:BVI} and ~\ref{fig:UBV}. The ordering of the various subpopulations in terms of $\Delta$(F336W$-$F438W) and $\Delta$(F336W$-$F814W)
is different from the case of optical colours, with the various P2 sub-populations moving to the red relative to P1 sub-populations, compared to optical colours. On the other hand,  
P1-b and P1-c (in NGC~2808) stars are still increasingly bluer than P1-a objects, as in optical CMDs.

This consistency of the RGB colours of P1 (and P2) stars with a range of helium abundances for the specific case of NGC~2808 agrees with the in-depth analysis by \cite{milone15}.
In their paper however, no link is made between the observed multiple He RGB sequences, and their belonging to low Na and high Na sub-populations.

\section{Discussion and Conclusions}\label{CONCLUSIONI}

It has been well established by \citet{milone17} that the distribution of first population stars in the chromosome maps of most --but not all--
clusters is more extended than expected from purely observational errors and its extension is strongly correlated with the cluster mass. So far no clear explanation for the reasons of the observed extended P1 distributions has been provided.
We have considered here three clusters
with similar metallicity ([Fe/H]$\sim -$1.3) and different chromosome maps (and masses), namely NGC~288, M~3 and NGC~2808, the latter two displaying an extended P1 made of different sub-populations.  
Using synthetic spectra calculations, we have shown that this can be explained if M~3 and NGC~2808 P1 stars contain a range of He and possibly N abundances.  
The He inhomogeneity of P1 stars in these clusters is also confirmed by comparing the colour distribution of the various P1 subpopulations in standard CMDs.
Among the nominal P1 stars there appears to be a significant He spread, $\Delta$Y$\sim$0.024 and $\sim$0.125 for M~3 and NGC~2808 respectively. 

We stress that the tests presented in \S~\ref{teoria} are not meant to directly estimate light element abundance variations for the clusters studied here, but rather provide an interpretation of the properties of the chromosome maps. Indeed, only a handful of GC stars have been fully characterised spectroscopically in terms of their abundances of C, N, O \citep[][]{review}  and more data is needed to perform a quantitative analysis. Moreover, UV theoretical spectra appear not reliable enough to directly estimate accurate abundance variations from photometry \citep[see ][for a complete discussion]{dotter15}. Nonetheless, we emphasise that while a quantitative characterisation of GC stars in terms of He, C, N, and O from their chromosome maps appear premature at this stage, the evidence that the range of colour spreads along the $x$- and $y$-axes in the chromosome maps correlate with the level of He and N enrichment is supported by the analysis presented in \S~\ref{CROMO}, that includes also a study of the P1 and P2 optical CMDs \citep[see also Table~4 in][]{milone15}.

As mentioned already in \S~\ref{teoria}, both Mg and Si variations amongst RGB P1 stars can affect their T$_{\rm eff}$ , 
mimicking the presence of He variations \citep{vdb}. 
However, the evidence of significant Mg and Si variations is found only for a few clusters \citep[e.g.][]{carretta2009,pancino17}, whereas the P1 groups display 
extended (F275W-F814W) colour sequences in the majority of clusters studied by \citet{milone17}. In the specific case of NGC~2808 investigated here (and in M~4 as discussed before), spectroscopic
analyses reveal that P1 stars are homogeneous in Mg and Si.


According to standard self-enrichment models for MPs, in a given cluster only P2 stars as defined by their chromosome map location 
 are expected to exhibit different levels of He and N, whereas P1 stars should have 
 essentially the same chemical composition. The presence of a range of initial He and N abundances amongst stars with homogeneous Na and O content, and the fact
 that these inhomogeneities 
appear only in some clusters, 
challenge all scenarios proposed so far to explain the MP phenomenon in globular clusters. 

Clusters that display extended P1 and P2 have two distinct chemical abundance patterns.  The upper branch -- starting around the origin of the chromosome maps and ending up with stars with high/low values of $\Delta$C(F275W, F336W, F438W)/$\Delta$(F275W-F814W)-- corresponds to the classical MP concept, with stars showing anti-correlated abundance patterns of Na-O, N-O and He-O.  While this branch still remains to be understood, it is well studied and has been confirmed to be present in nearly all the ancient GCs. The lower branch, however, is not universal amongst the ancient clusters.  Some clusters have a tightly defined P1 population without any signs of internal abundance variations, while others display extended (and chemically inhomogeneous) P1 populations.  

The extended P1 populations appear to be homogeneous in O, and instead are mainly driven (in the chromosome maps) by a spread in He abundances with potentially a small variations in N as well.  Such a pattern cannot be explained through the hot hydrogen burning channels that have generally been adopted to explain the classical multiple populations (i.e. P2) as large spreads in C, N, O, Na and Al would be expected.  Instead, the extended P1 pattern is more reminiscent of P-P chain hydrogen burning.  This wholly unexpected results adds to the enigma of the multiple populations phenomenon.

\begin{acknowledgements}
We would like to warmly thank S.~Cassisi, S.~Larsen, S.~Martell, E.~Pancino, and M.~Rejkuba for useful discussion. We thank the anonymous referee for meritorious suggestions. CL acknowledges financial support from the Swiss National Science Foundation (Ambizione grant PZ00P2\_168065). NB gratefully acknowledges financial support from the Royal Society (University Research Fellowship) and 
the European Research Council (ERC-CoG-646928, Multi-Pop). Support for this work was provided by NASA through Hubble Fellowship grant HST-HF2-51387.001-A awarded by the Space Telescope Science Institute, which is operated by the Association of Universities for Research in Astronomy, Inc., for NASA, under contract NAS5-26555.
\end{acknowledgements}

%
%

\bibliographystyle{aa} 
 \bibliography{bibliography}

\end{document}